\documentclass[12pt]{iopart}

\usepackage{iopams}
\usepackage{multirow}
\usepackage{graphicx}
\usepackage{mathabx}
\usepackage{bbm}
\usepackage{amssymb}
\usepackage{booktabs}
\usepackage{appendix} 
\usepackage{lscape}

\begin{document}

\title{Can the coincidence problem be solved by a cosmological model of coupled dark energy and dark matter?}
\author{Vincent Poitras}
\address{McGill University,\\
3600 University street, Montreal, Canada}
\eads{\mailto{poitrasv@physics.mcgill.ca}}

\begin{abstract}
{
Motivated by the cosmological constant and the coincidence problems, we consider a cosmological model  where the dark sectors are interacting together through a phenomenological decay law $\dot{\rho}_{\rm \Lambda}=Q\rho_{\rm \Lambda}^n$ in a FRW spacetime with spatial curvature. We show that the only value of $n$ for which the late-time matter energy density to dark energy density ratio ($r{\rm _m}=\rho_{\rm m}/\rho_{\rm \Lambda}$) is constant (which could provide an explanation to the coincidence problem) is $n=3/2$. For each value of $Q$, there are two distinct solutions. One of them involves a spatial curvature approaching zero at late times ($\rho_{\rm k}\approx0$) and is stable when the interaction is weaker than a critical value ${Q_0=-\sqrt{32\pi G/c^2}}$. The other one allows for a non-negligible spatial curvature ($\rho_{\rm k}\napprox0$) at late times and is stable when the interaction is stronger than $Q_0$. We constrain the model parameters using various observational data (SNeIa, GRB, CMB, BAO, OHD). The limits obtained on the  parameters exclude the regions where the cosmological constant problem is significantly ameliorated and do not allow for a completely satisfying explanation  for the coincidence problem.
}
\end{abstract}
\submitto{\CQG}

  \maketitle 
  \flushbottom

\section{Introduction}
\label{sect:Intro}

It is now more than a decade since the first observations of type Ia supernovae suggesting that the Universe is currently experiencing a phase of accelerated expansion were done \cite{SNE1,SNE2,SNE3}. Since then, improved measurement of supernovae distance \cite{SNE4,SNE5} and additional evidence based, for instance, on the measurement of the cosmic microwave background \cite{CMB1,CMB2,CMB3,CMB4,planck} or on the apparent size of the baryons acoustic oscillations \cite{BAO1,BAO2} have led to the same conclusion. The $\Lambda$CDM model is currently considered to be the most successful cosmological model by reason of its simplicity and of the quality of the fit to the data that it provides. In this model, the Universe is composed of, in addition to ordinary matter (radiation, baryon), a pressureless cold dark matter fluid and a cosmological constant $\Lambda$, the simplest form  of dark energy. 

However, despite the excellent agreement with the observational data, the $\Lambda$CDM model is facing two theoretical difficulties, namely the \emph{cosmological constant problem} \cite{Weinberg} and the \emph{coincidence problem} \cite{CoinProb}. Regarding the first one, there is a discrepancy of $\sim$123  orders of magnitude between the value of the energy density expected from theoretic computation and the value inferred from observations ($\rho_{\Lambda_{\rm{obs}}}/\rho_{\Lambda_{\rm{th}}}\sim10^{-123}$). As for the second one, according to the observations, the current values of the energy densities of matter and of dark energy are of the same order of magnitude ($\rho_{\rm m_0}/\rho_{\Lambda_0}=\Or(1)$). This is not strictly incompatible with the model, but however, requires a fine tuning of the initial conditions of the model.

A possible way to avoid these problems would be to replace the cosmological constant $\Lambda$ by a cosmological term, $\Lambda(t)$, which is allowed to vary in time (see for instance \cite{cstr0,cstr1,cstr2,cstr4,cstr8,cstrA1,cstr3,cstr5,cstr6,cstr9,intmod2,intmod1} for some recent examples and \cite{rev1,rev2,rev3} for a  review). Hence, it would be possible for the dark energy density to decrease from an initial large value, consistent with the theoric expectation, to a smaller one, consistent with the current value inferred from the observations. In \cite{LtCDM}, we proposed a phenomenological model, referred to the $\Lambda(t)$CDM model, where the dark fluids are interacting together in a flat spacetime with an energy transfer rate of the form $Q_{\rm \Lambda}\propto\rho_{\rm \Lambda}^n$. We  mainly focused on the case where $n=3/2$ since, as we showed, it is the only one for which the ratio of  matter to dark energy densities remains constant at late times ($r_{\rm m}\equiv\rho_{\rm m}/\rho_{\rm \Lambda}=const$). This could have provided an explanation for the coincidence problem since the current value of $r_{\rm m}$ could thus become typical of late times; however it turned out that the region of the parameter space where the coincidence and the cosmological constant problems are solved (or at least significantly alleviated) are excluded by the observational constraints.

The aim of this paper is to extend the analysis of \cite{LtCDM} for a flat spacetime to one with spatial curvature. 
Moreover, in our previous work, we had set the value of the energy density of radiation to that obtained in the context of the $\Lambda$CDM model in order to reduce the number of parameters to constrain. Here, we will consider this quantity as a free parameter. 

\section{Models}
\label{sect:IDS}
\subsection{Basic equations}

In a Friedmann-Robertson-Walker (FRW) spacetime, if  the Universe content is modeled by perfect fluids, its continuity equation is given by
  \begin{equation}
     \dot{\rho} = -3H(\rho+p),
     \label{eq:contall}
  \end{equation}
where $H\equiv\dot{a}/a$ ($a$ is the scale factor) stands for the Hubble term, $\rho$ is the sum of the energy density of each fluid ($\rho=\sum_i \rho_i$) and similarly, $p$ is the sum of the pressure of each fluid ($p=\sum_i p_i$). Defining $Q_i\equiv\dot{\rho}_i+3H(\rho_i+p_i)$, we obtain a continuity equation for each fluid, subject to the the condition $\sum_{i}Q_{i}=0$. These equations can be more conveniently written as
\begin{equation}
     \dot{\rho}_i = -3H(1+w_i)\rho_i + Q_{i},
     \label{eq:cont}
\end{equation}
where the value of the equation of state (EoS) parameter ($w_i\equiv p_i/\rho_i$) depends on the nature of the fluid ($w_{\rm m}=0$ for cold matter (dark and baryonic), $w_{\rm r}=1/3$ for radiation and $w_{\rm \Lambda}=-1$ for dark energy). As we can see from the previous equation, the variation of the energy density could be the result of two different mechanisms. The first term on the RHS represents the usual  energy density dilution caused by the cosmic expansion. As for the other term, since $\sum_{i}Q_{i}=0$, it must be interpreted as  a possible energy transfer between the fluids. A positive value ($Q_i>0$) constitutes  a gain of energy for the fluid (source term), and negative value ($Q_i<0$),  a loss of energy (sink term). In the $\Lambda$CDM model, the energy of each fluid is conserved separately, i.e. $Q_i=0$ for all of them. 

In addition to (\ref{eq:cont}), to completely specify the time evolution we also need the Friedmann equation, which takes its usual form\footnote[1]{In order to have a more compact notation, we treat here the contribution of spatial curvature as a fictitious fluid whose energy density is defined as $\rho_{\rm k}\equiv-3\kappa c^4/8\pi G a^2$. The curvature parameter $\kappa$, whose dimensions are (length)$^{-2}$, is negative for an open Universe and positive for a closed one. In (\ref{eq:cont}), $w_{\rm k}=-1/3$ and $Q_{\rm k}=0$. A non-zero value for $Q_{\rm k}$ would be inconsistent with the FRW metric.}
    \begin{equation}
      H^2=\frac{8\pi G}{3c^2}(\rho+\rho_{\rm k}).
     \label{eq:fried1}
  \end{equation}

Now it remains only to specify the interaction terms $Q_i$ for the $\Lambda(t)$CDM model. Following our previous work \cite{LtCDM}, we will chose the interaction term between dark energy and dark matter ($Q_{\rm dm}=-Q_{\rm \Lambda}$) to be $Q_{\rm \Lambda}=Q\rho_{\rm \Lambda}^{n}$, where $Q$ is a parameter to constrain. In order to find an explanation to the coincidence problem, we will try to find under which conditions, if any,  it is possible to obtain a phase during which the ratio of the dark matter to the dark energy densities ($r_{\rm m}\equiv \rho_{\rm m}/\rho_{\rm \Lambda}$) remains constant ($\dot{r}_{\rm m}=0$). To simplify our analysis, we will first consider the case of an era dominated by dark energy and matter ($\Lambda$m-dominated era, $\rho_{\rm \Lambda},\rho_{\rm m}\gg\rho_{\rm k},\rho_{\rm r}$), and subsequently, that of an era dominated by dark energy, matter and curvature ($\Lambda$mk-dominated era, $\rho_{\rm \Lambda},\rho_{\rm m},\rho_{\rm k}\gg\rho_{\rm r}$).

\subsection{$\dot{r}_{\rm m}=0$ during a $\Lambda$m-dominated era}
\label{subsect:lm-era}
In \cite{LtCDM}, we have already shown that is impossible to have a constant ratio $r_{\rm m}$  during $\Lambda$m-dominated era unless that $n=3/2$. Indeed, using the continuity equations  for dark energy and matter (\ref{eq:cont}), we can show that
\begin{equation}
     \dot{r}_{\rm m}=-3Hr_{\rm m}-Q(1+r_{\rm m})\rho_{\rm \Lambda}^{n-1}.
     \label{eq:ratio}
 \end{equation}
If we suppose that at some point, $r_{\rm m}$ reaches a constant value $\tilde{r}_{\rm m}$, hence $\dot{r}_{\rm m}=0$ at this point, and the parameter $Q$ will be related to this value through 
\begin{equation}
     Q=-3\left(\frac{\tilde{r}_{\rm m}}{1+\tilde{r}_{\rm m}}\right)H\rho_{\rm \Lambda}^{1-n}.
     \label{eq:Qrdot0}
 \end{equation}
Since $Q$ is a constant,  the product $H\rho_{\rm \Lambda}^{1-n}$ must also be a constant. During a $\Lambda$m-era, the Hubble term may be written as
  \begin{equation}
     H= \pm\left[\frac{8\pi G}{3c^2}(1+r_{\rm m})\right]^{\frac{1}{2}}\rho_{\rm \Lambda}^{\frac{1}{2}}.
     \label{eq:Hlm}
  \end{equation}
The plus sign  stands for an expanding Universe and the minus sign for a contracting one. We will only consider the former case ($H>0$).  Inserting this expression into (\ref{eq:Qrdot0}), we see that the only consistent value for $n$ is $3/2$, which leads to
\begin{equation}
     Q=-\sqrt{\frac{24\pi G}{c^2}}\frac{\tilde{r}_{\rm m}}{(1+\tilde{r}_{\rm m})^{\frac{1}{2}}}.
     \label{eq:QLm}
 \end{equation}
The negative value for the interaction term implies that the energy transfer must occur from dark energy to dark matter in order to reach a phase with a constant ratio $\tilde{r}_{\rm m}$ in an expanding Universe. If we invert this equation, we finally get an expression for $\tilde{r}_{\rm m}$
\begin{equation}
     \tilde{r}_{\rm m}=\frac{Q^2}{48\pi G/c^2}\left(1+\sqrt{1+\frac{96\pi G/c^2}{Q^2}}\right),
     \label{eq:rm}
 \end{equation}
which is shown in figure~\ref{fig:qq}.

In \cite{LtCDM}, we have shown that, provided that $Q<0$, a flat Universe will necessarily experience a late phase with $\dot{r}_{\rm m}=0$. For a non-flat Universe, this result does not necessarily hold since the energy associated with the spatial curvature could possibly become non-negligible before that this phase has been reached. Thus, we have to determine if $\rho_{\rm k}$  decreases slower or faster than $\rho_{\rm \Lambda}$ and $\rho_{\rm m}$. In the $\Lambda$CDM model, to answer this question, we simply have to compare the EoS parameter of the fluids; a smaller value implies that the energy density will decrease slower. In the $\Lambda(t)$CDM model, to take in account the effect of the energy transfer between the fluids, we need to look at the effective EoS parameter,  defined as
$w_i^{{\rm eff}}\equiv w_i-\frac{Q_i}{3H\rho_i}$. The continuity equation now takes the same form as in the $\Lambda$CDM model
\begin{equation}
     \dot{\rho}_i = -3H(1+w_i^{{\rm eff}})\rho_i.
     \label{eq:conteff}
\end{equation}
Since $Q_{\rm k}=0$, $w_{\rm k}^{{\rm eff}}=-1/3$ for the curvature. For dark energy and matter, in the case where $r_{\rm m}$ is approaching $\tilde{r}_{\rm m}$, the effective EoS parameters become $w_{\rm \Lambda}^{{\rm eff}}=w_{\rm m}^{{\rm eff}}=-1/(1+\tilde{r}_{\rm m})$. Therefore,  the energy density $\rho_{\rm k}$ will decrease faster than the two other (${w_{\rm k}^{{\rm eff}}>w_{\rm \Lambda}^{{\rm eff}}=w_{\rm m}^{{\rm eff}}}$)  if $\tilde{r}_{\rm m}<2$ (or equivalently, if $\tilde{\Omega}_{\rm \Lambda}>1/3$). In this case the approximation of a $\Lambda$m-dominated era will remain accurate for ever. Conversely, $\rho_{\rm k}$ will decrease slower if $\tilde{r}_{\rm m}>2$ ($\tilde{\Omega}_{\rm \Lambda}<1/3$). In this case, it would be possible to find solutions where $r_{\rm m}$ approaches $\tilde{r}_{\rm m}$ for a certain time, but eventually the assumption of a $\Lambda$m-dominated era will become invalid. We then need to look what happen during a  $\Lambda$mk-dominated era.

\subsection{$\dot{r}_{\rm m}=0$ and $\dot{r}_{\rm k}=0$ during a $\Lambda$mk-dominated era}
\label{subsect:lmk-era}

\begin{figure}[!ht]
  \centering
  \includegraphics[scale=0.650]{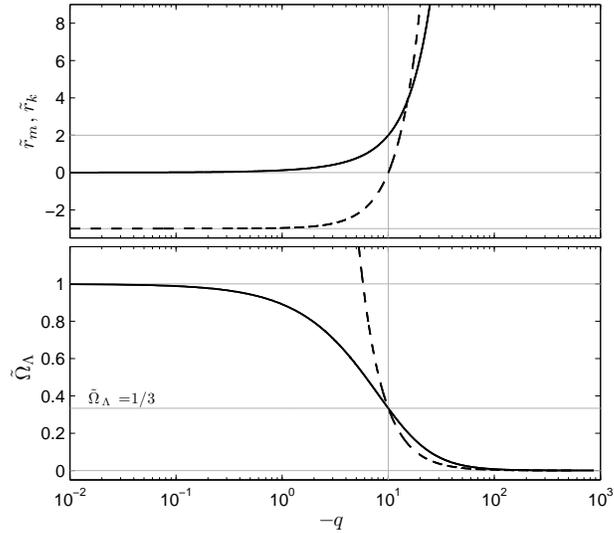} 
  \caption{ In the upper panel, the functions $\tilde{r}_{\rm m}$ (solid line) and  $\tilde{r}_{\rm k}$ (dashed line) are shown as a function of the dimensionless parameter $q\equiv Q(c/G^{1/2})$. In the lower panel, the density parameter of dark energy ${\Omega_{\rm \Lambda}=(1+r_{\rm m}+r_{\rm k})^{-1}}$ is shown for the solution found in section \ref{subsect:lm-era}, $(r_{\rm m},r_{\rm k})=(\tilde{r}_{\rm m},0)\rightarrow{\tilde{\Omega}_{\rm \Lambda}=(1+\tilde{r}_{\rm m})^{-1}}$, (full line) and for the solution found in section~\ref{subsect:lmk-era}, $(r_{\rm m},r_{\rm k})=(2,\tilde{r}_{\rm k})\rightarrow{\tilde{\Omega}_{\rm \Lambda}=(3+\tilde{r}_{\rm k})^{-1}}$, (dashed line). In both panels, the vertical line corresponds to the point ${Q_0=-\sqrt{32\pi G/c^2}}$ (${q_0=-\sqrt{32\pi}}\approx-10$) where the two solutions are equivalent $(r_{\rm m},r_{\rm k})=(\tilde{r}_{\rm m},\tilde{r}_{\rm k})=(2,0)$.}
  \label{fig:qq}
\end{figure}

During a $\Lambda$mk-dominated era, the Hubble term may be written as
  \begin{equation}
     H= \pm\left[\frac{8\pi G}{3c^2}(1+r_{\rm m}+r_{\rm k})\right]^{\frac{1}{2}}\rho_{\rm \Lambda}^{\frac{1}{2}}.
     \label{eq:Hlmk}
  \end{equation}
As in the previous section, we will consider only the case of an expanding Universe ($H>0$). Moreover, the product $H\rho_{\rm \Lambda}^{1-n}$ must still be a constant in order to have $\dot{r}_{\rm m}=0$ (c.f. (\ref{eq:Qrdot0})). But now, this product involves the curvature to dark energy density ratio ($r_{\rm k}\equiv\rho_{\rm k}/\rho_{\rm \Lambda}$) and is proportional to $\sqrt{1+r_{\rm m}+r_{\rm k}}\rho_{\rm \Lambda}^{3/2-n}$. For $n=3/2$, this quantity will be a constant only if, in addition to the ratio $r_{\rm m}$, the ratio $r_{\rm k}$ is also a constant. In this case, it is useful to derive from (\ref{eq:cont}) an equation for the time derivative of $r_{\rm k}$  
\begin{equation}
     \dot{r}_{\rm k}=-2Hr_{\rm m}-Qr_{\rm k}\rho_{\rm \Lambda}^{n-1}.
     \label{eq:rkdot}
 \end{equation}
Setting $\dot{r}_{\rm k}=0$ ($r_{\rm k}=\tilde{r}_{\rm k}$), we obtain an equation analogous to (\ref{eq:Qrdot0})
\begin{equation}
     Q=-2H\rho_{\rm \Lambda}^{1-n}.
     \label{eq:Qrdotk}
 \end{equation}
These two expressions for the parameter $Q$ must be equivalent and that will be the case only if $r_{\rm m}=2$, which leads to
\begin{equation}
     Q=-\frac{2}{3}\sqrt{\frac{24\pi G}{c^2}}(\tilde{r}_{\rm k}+3)^{\frac{1}{2}}.
     \label{eq:QLmk}
 \end{equation}
As in (\ref{eq:QLm}), the interaction term is negative. Inverting this equation, we get
\begin{equation}
     \tilde{r}_{\rm k}=3\left(\frac{Q^2}{32\pi G/c^2}-1\right).
     \label{eq:rk}
 \end{equation}
This function is shown in figure~\ref{fig:qq}. In this figure, one sees that for each value of $Q$, there are actually two solutions for which $\dot{r}_{\rm m}=0$ (corresponding to $\tilde{r}_{\rm m}$ and  $\tilde{r}_{\rm k}$). Hence, we have to find under which conditions one or the other solution (if any) will be relevant. First of all, we can notice that the two solutions are equivalent when ${(r_{\rm m},r_{\rm k})=(\tilde{r}_{\rm m},\tilde{r}_{\rm k})=(2,0)}$  (which corresponds to $\tilde{\Omega}_{\rm \Lambda}=1/3$). In this case, the interaction parameter is  given by ${Q_0\equiv-\sqrt{32\pi G/c^2}}$. It turns out that the cosmic evolution will be qualitatively different depending on whether the strength of interaction is weaker ($|Q|<|Q_0|$) or stronger ($|Q|>|Q_0|$) than this critical value.

\begin{figure}[!ht]  
    \centering
    \begin{tabular}{ll}
     \includegraphics[scale=0.650]{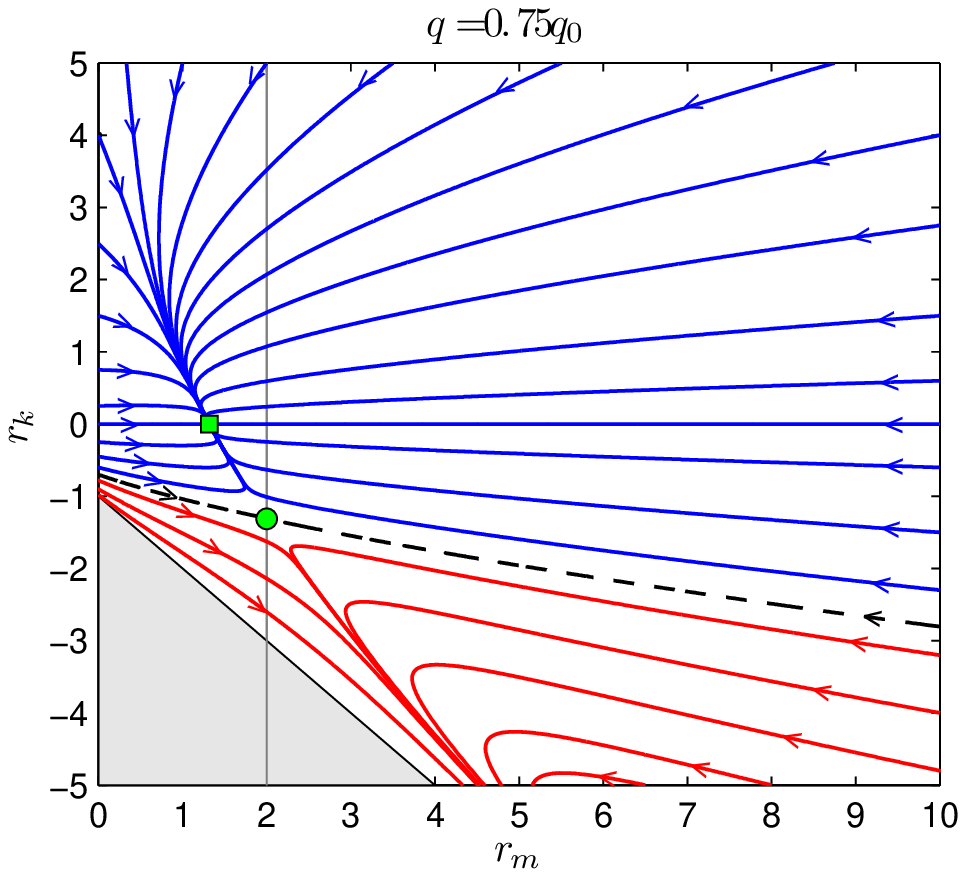} & \includegraphics[scale=0.650]{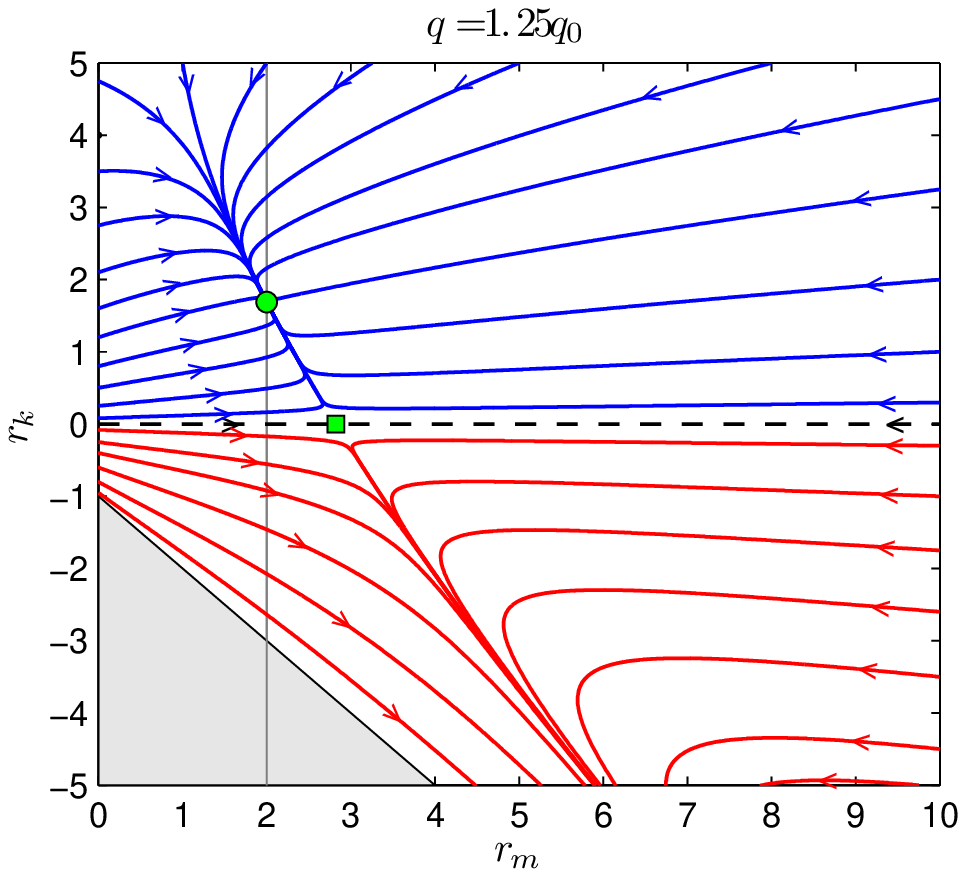}
    \end{tabular}  
  \caption{Examples of trajectories in the plane $r_{\rm m}-r_{\rm k}$ for an expanding Universe ($H>0$) where the radiation is neglected  ($\rho_{\rm r}=0$). In the left panel, the interaction parameter is smaller (in magnitude) than the critical value $q_0$ ($q=0.75q_0$), and in the right panel,  it is larger (in magnitude) than $q_0$ ($q=1.25q_0$).  The solution  found in section~\ref{subsect:lm-era} for which $\dot{r}_{\rm m}=0$ ($\tilde{r}_{\rm k}=0$, $\tilde{r}_{\rm m}$ given by (\ref{eq:QLm}) is represented by a square mark, and that found in section~\ref{subsect:lmk-era} ($\tilde{r}_{\rm m}=2$, $\tilde{r}_{\rm k}$ given by (\ref{eq:QLmk}), by a circular mark. The positive values of $r_{\rm k}$ correspond to a negatively curved space (open Universe) and the negative ones to a positively curved space (closed Universe). The area under the line $r_{\rm k}=-(r_{\rm m}+1)$ corresponds to a non-physical region where $H^2<0$.}
  \label{fig:attrak}
\end{figure}

In  figure~\ref{fig:attrak},  two examples of trajectories in the plane ${r}_{\rm m}-{r}_{\rm k}$ are shown, one where the interaction is weaker than $Q_0$ and the other where it is stronger. In each case, the plane is divided into two regions. In the figure,  the boundary between them is represented by a dashed line.  For the region situated above this line, all the trajectories end at the same point. For the weak case, this point corresponds to the flat solution found in the previous section ($(r_{\rm m},r_{\rm k})=(\tilde{r}_{\rm m},0)$) and for the strong case, to the non-flat solution found in the current section ($(r_{\rm m},r_{\rm k})=(2,\tilde{r}_{\rm k})$). In the region situated below the dashed line, the fate of the Universe will be the same no matter the value of $Q$; it will eventually reach a point on the line $r_{\rm k}=-(r_{\rm m}+1)$ (where $H=0$) and then recollapse. As for the trajectories starting exactly on the boundary between these two regions, they will also end in a point  determined only by the value of $Q$, but conversely to the upper region, this point corresponds now to the non-flat solution for the weak  case and to the flat solution for strong  case. However, as we can see from the figure, these solutions are unstable since any small perturbation which takes the trajectory slightly away from the dashed line  will make it diverge toward the stable solutions in the upper region or toward a recollapsing point in the lower region. For the strong  case, the boundary between the two regions is set by the line $r_{\rm k}=0$, while for the weak  case, the boundary is entirely situated below this line (in this case we cannot obtain an analytic expression to describe it). Thus, an open Universe ($r_{\rm k}>0$) will always evolve up to reach a stable point, while for a closed Universe ($r_{\rm k}<0$), that will be possible for a given initial point ($r_{\rm m_0},r_{\rm k_0}$) only if the interaction is sufficiently weak (at least $|Q|<Q_0$).

Here we have to keep in mind that in order to be able to explain the coincidence problem, it is not sufficient to find a solution for which the ratio $r_{\rm m}$ becomes constant; the current value of the ratio ($r_{\rm m_0}$) must also be close to this constant value . For the non-flat solution, this means that this value should be close to two ($r_{\rm m_0}\approx2$). This value is so different from that obtained for the $\Lambda$CDM model (which already provides a good fit to data) that it is reasonable to expect that this solution (and the larger values of $|Q|$ associated with it)  will be excluded by the observational constraints. If the larger values of $|Q|$ are excluded, that will also affect the ability of the $\Lambda(t)$CDM model to solve the cosmological constant problem. Indeed, if the interaction is too weak, that will not be possible for the dark energy density to decay from an initial large value to the small one observed today. Actually, the condition $r_{\rm m}=2$ for the non-flat solution results from the fact that, for $n=3/2$, in  order to have a constant value of $r_{\rm m}$, $r_{\rm k}$ must also be a constant (cf. (\ref{eq:Qrdotk})). It would then be interesting to verify whether it is possible to obtain a solution with $\dot{r}_{\rm m}=0$ and $\dot{r}_{\rm k}\neq0$ during a $\Lambda$mk-dominated era if we consider a different value of $n$. In other words, we would like to check if it is possible to find a solution to $H\rho_{\rm \Lambda}^{1-n}\propto\sqrt{1+\tilde{r}_{\rm m}+r_{\rm k}(t)}\rho_{\rm \Lambda}^{3/2-n}(t)=const$ for $n\neq3/2$.

\subsection{$\dot{r}_{\rm m}=0$ and $\dot{r}_{\rm k}\neq0$ during a $\Lambda$mk-dominated era?}

To derive (\ref{eq:Qrdot0}), which implicitly implies that $\dot{r}_{\rm m}=0$, we have used the continuity equations of  dark energy and of  matter. During a $\Lambda$mk-dominated era, the description of the Universe also involved the continuity equation of curvature and the Friedmann equation. We can use these two equations to check whether  there are other values than $n=3/2$ that are consistent with the condition $\dot{r}_{\rm m}=0$. From the Friedmann equation (\ref{eq:Hlmk}), we get
  \begin{equation}
     \rho_{\rm k}=\frac{3c^2}{8\pi G}H^2-(1+r_{\rm m})\rho_{\rm \Lambda}.
     \label{eq:rhok}
  \end{equation}
In order to have $\dot{r}_{\rm m}=0$, the Hubble term must be given  (\ref{eq:Qrdot0}), i.e. 
  \begin{equation}
     H=-\frac{(1+\tilde{r}_{\rm m})Q}{3\tilde{r}_{\rm m}}\rho_{\rm \Lambda}^{n-1}.
     \label{eq:hn}
  \end{equation}
Hence, differentiating (\ref{eq:rhok}) and replacing $r_{\rm m}$ by $\tilde{r}_{\rm m}$ yields
  \begin{equation}
     \dot{\rho}_{\rm k}=\left[\frac{(n-1)c^2}{12\pi G}\left(\frac{1+\tilde{r}_{\rm m}}{\tilde{r}_{\rm m}}\right)^2 Q^3\right]\rho_{\rm \Lambda}^{3n-3}-\left[(1+\tilde{r}_{\rm m})Q\right]\rho_{\rm \Lambda}^n.
     \label{eq:rhokd1}
  \end{equation}
This expression has to be compared to the continuity equation of curvature (\ref{eq:cont}), which becomes, using the expression of $\rho_{\rm k}$ and $H$ given by (\ref{eq:rhok}) and (\ref{eq:hn})
 \begin{equation}
     \dot{\rho}_{\rm k}=\left[\frac{c^2}{36\pi G}\left(\frac{1+\tilde{r}_{\rm m}}{\tilde{r}_{\rm m}}\right)^3 Q^3\right]\rho_{\rm \Lambda}^{3n-3}-\left[\frac{2}{3}\left(\frac{1+\tilde{r}_{\rm m}}{\tilde{r}_{\rm m}}\right)(1+\tilde{r}_{\rm m})Q\right]\rho_{\rm \Lambda}^n.
     \label{eq:rhokd2}
  \end{equation}
These two expressions for $\dot{\rho}_{\rm k}$ are equivalent in two cases: ($\tilde{r}_{\rm m}=2$, $n=3/2$) and ($\tilde{r}_{\rm m}=-1$, $n$ unfixed). We have already considered the first one in the previous section. The second one must be rejected, since, in addition to involve a violation of the weak energy condition ($\rho\geq0$) either for $\rho_{\rm \Lambda}$ or for $\rho_{\rm m}$, this solution has been obtained from a division by zero in (\ref{eq:Qrdot0}). Hence, we conclude that the only value leading to  $\dot{r}_{\rm m}=0$ during a $\Lambda$mk-dominated era is $n=3/2$. Consequently, it is impossible to have simultaneously $\dot{r}_{\rm m}=0$ and  $\dot{r}_{\rm k}\neq0$.

\section{Results and discussion}
\label{sect:RnD}
To assess the validity of the $\Lambda(t)$CDM model, we have constrained the model parameters using the methodology described in appendix A. For the $\Lambda$CDM model, the continuity equations (\ref{eq:cont}) of the four fluids (dark energy, matter, radiation and curvature) involve five parameters. They can be chosen as $\rho_{\Lambda_0}$, $\rho_{\rm m_0}$, $\rho_{\rm r_0}$, $\rho_{\rm k_0}$ and $H_0$, where as usual, the subscript zero refers to the current value of these quantities. However, due to  the Friedmann equation, only four of them are independent. For the $\Lambda(t)$CDM model we have in addition to consider the interaction parameter $Q$. It will be more convenient, but completely equivalent, to express our results in terms of the following dimensionless parameters
  \begin{equation}
   \Omega_{i}\equiv\frac{\rho_i}{\rho+\rho_{\rm k}}, \hspace{25 pt} h \equiv \frac{H}{100\ \rm{km\  s^{-1}\ Mpc^{-1}}}, \hspace{25 pt} q\equiv\frac{c}{G^{\frac{1}{2}}}Q.
  \end{equation}
Concretely, we have constrained the following five parameters: $\Omega_{\Lambda_{0}}$, $\Omega_{m_{0}}$, $\Omega_{r_{0}}$, $h_0$ and $q$. The current value of the density  parameter of curvature, $\Omega_{\rm k_0}$, may be obtained from the relation $\Omega+\Omega_{\rm k}=1$, where ${\Omega\equiv\sum_{i\neq {\rm k}}\Omega_i}$.

\begin{figure}[!ht]
  \centering
  \includegraphics[scale=0.700]{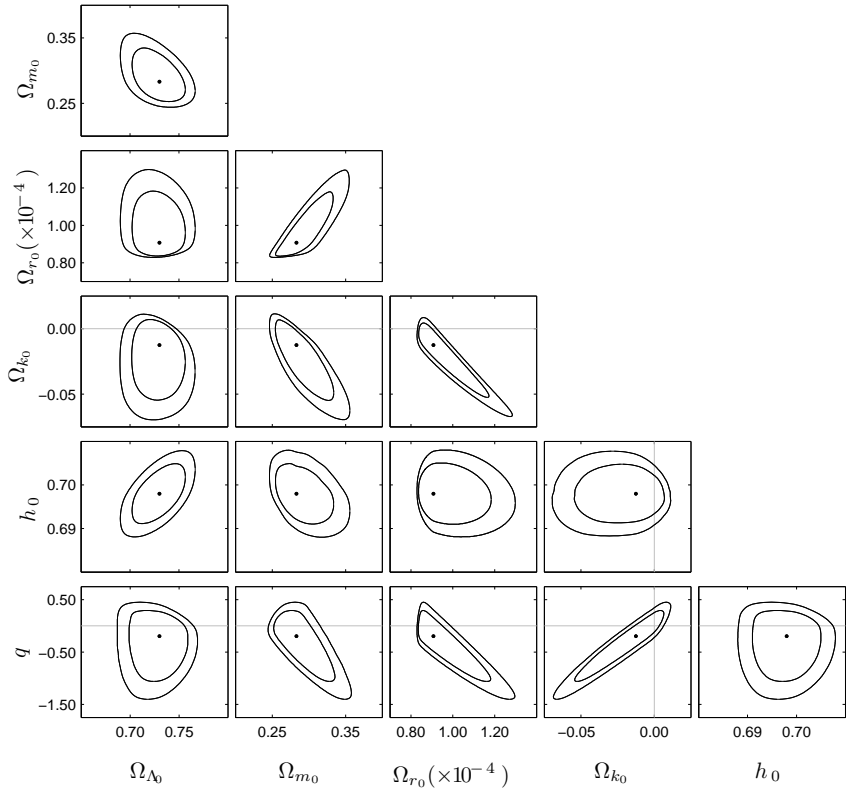} 
  \caption{Projections  of the 1-$\sigma$ and 2-$\sigma$ confidence regions obtained from observational constraints (c.f.  appendix~A) in the planes formed by the combination of two of the following parameters: $\Omega_{\Lambda_{0}}$, $\Omega_{m_{0}}$, $\Omega_{r_{0}}$, $\Omega_{k_{0}}$, $h_0$ and $q$. The best-fit values are also indicated by a dot in each panel. The energy density parameter of curvature was computed using the relation ${\Omega_{\rm k}=1-\Omega}$.  A negative value  ($\Omega_{\rm k_0}<0$) corresponds to a positive spatial curvature (closed Universe) and a positive value ($\Omega_{\rm k_0}>0$), to a negative space curvature (open Universe). In the panels where the interaction parameter $q$ is involved, the negative values ($q<0$) correspond to the decay of dark energy into dark matter and the positive values ($q>0$), to the inverse process. }
   \label{fig:g15}
\end{figure}

\begin{figure}[!ht]
  \centering
  \includegraphics[scale=0.800]{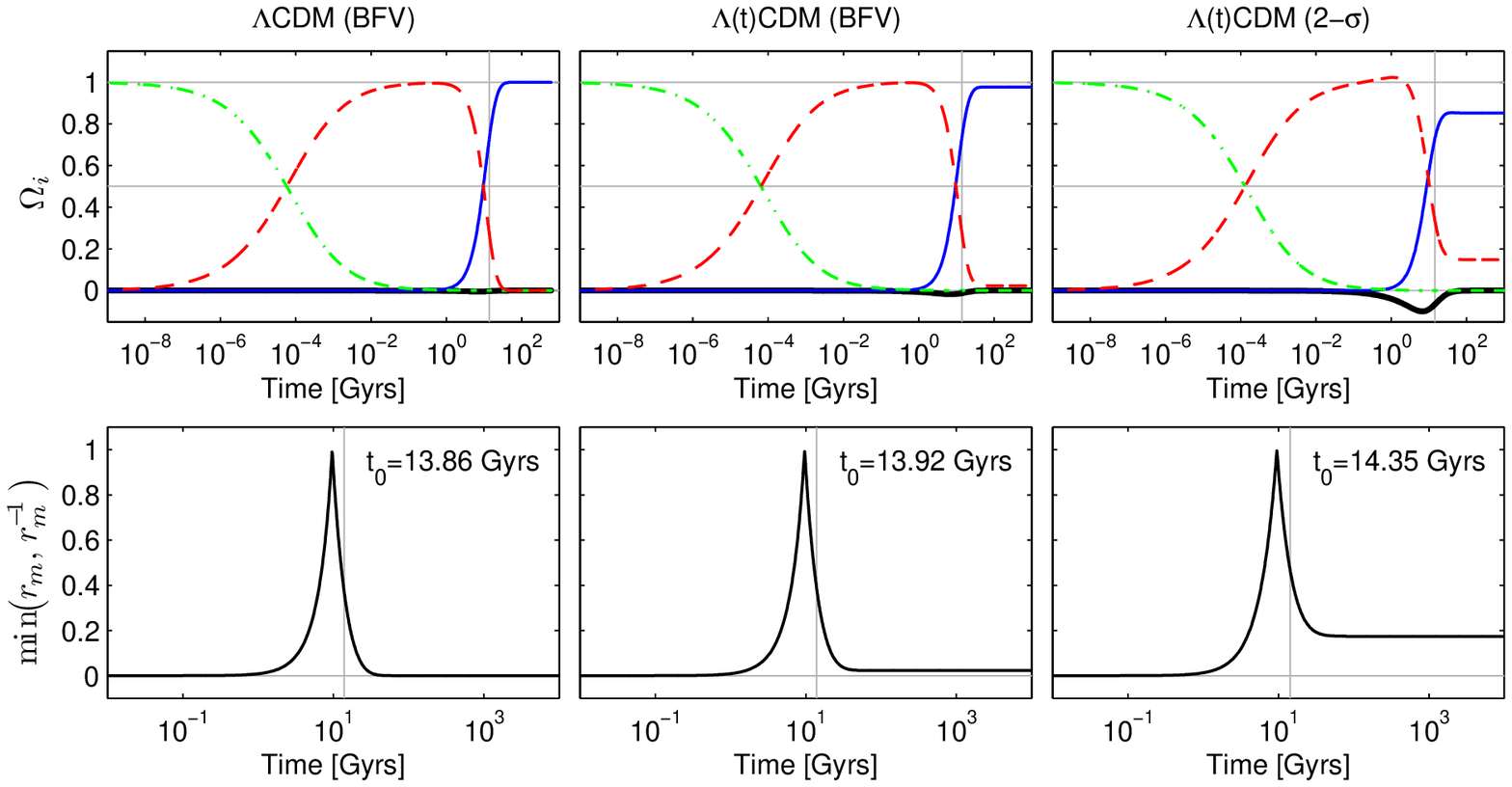} 
  \caption{ Upper row: evolution of the energy density parameter of dark energy (solid thin, blue), matter (dashed, red), radiation (dot-dashed, green) and  curvature (solid thick, black) as a function of the time.  Lower row:  $\min\left(r_{\rm m},r_{\rm m}^{-1}\right)$ as a function of the time. In each figure, the vertical line represents the current time $t_0$ and its numerical value is indicated in the lower row. The  first two columns correspond respectively to the best-fit values of the $\Lambda$CDM model and of the $\Lambda(t)$CDM (c.f. table~\ref{tab:bfv}) and the last row to the value of the parameters minimizing the quantity $\Delta{r_{\rm m}}/r_{\rm m_0}$ in the 2-$\sigma$ confidence region (c.f. figure~\ref{fig:grm}). }
  \label{fig:evol}
\end{figure}

The value of each parameter at the best-fit point and the corresponding $\chi_{\rm{min}}^2$ are shown for both models in table~\ref{tab:bfv}. The limits are the  extremal values of the 1-$\sigma$ and the 2-$\sigma$ confidence regions and they are shown  in figure~\ref{fig:g15} for the $\Lambda(t)$CDM model. At the best-fit point, the results that we obtained for the $\Lambda(t)$CDM model are not too much different from those of the $\Lambda$CDM model. Indeed, the values of the best-fit parameters of the $\Lambda$CDM model are all included in the 1-$\sigma$ confidence region of the $\Lambda(t)$CDM and, except for the interaction parameter $q$, the converse is also true. Moreover, in figure~\ref{fig:evol}, we can see that the evolution history of each fluid is relatively similar for each model.

\begin{scriptsize}
hello
\end{scriptsize}

\begin{table}

\caption{\label{tab:bfv}Best-fit values for the free parameters  ($\Omega_{\Lambda_{0}}$, $\Omega_{m_{0}}$, $\Omega_{r_{0}}$, $h_0$ and $q$) and the corresponding $\chi_{\rm min}^{2}$ for the $\Lambda$CDM  and the $\Lambda(t)$CDM models. The value for $\Omega_{k_{0}}$ has been computed using the relation ${\Omega_{\rm k}=1-\Omega}$.  The limits are the  extremal values of the 1-$\sigma$ and the 2-$\sigma$ confidence regions (shown  in figure~\ref{fig:g15} for the $\Lambda(t)$CDM model).}
\begin{indented}
\item[]\begin{tabular}{lcccc}

\br
  {\scriptsize                                    }& 
  {\scriptsize $\Omega_{\Lambda_0}$               }& 
  {\scriptsize $\Omega_{m_{0}}$                   }& 
  {\scriptsize $\Omega_{r_{0}}$ ($\times10^{-5}$) }& 
  {\scriptsize $\Omega_{k_{0}}$ ($\times10^{-2}$) }\\
\mr
{\scriptsize $\Lambda$CDM                                            }& 
{\scriptsize    0.729 $ ^{+0.024 }_{-0.027 }$ $ ^{+0.034 }_{-0.040 }$}& 
{\scriptsize    0.274 $ ^{+0.027 }_{-0.020 }$ $ ^{+0.039 }_{-0.028 }$}& 
{\scriptsize    8.55  $ ^{+0.87  }_{-0.19  }$ $ ^{+1.27  }_{-0.26  }$}&   
{\scriptsize $-$0.281 $ ^{+00.637}_{-01.009}$ $ ^{+00.912}_{-01.457}$}\\

{\scriptsize $\Lambda(t)$CDM                                         }& 
{\scriptsize    0.730 $ ^{+0.029 }_{-0.031 }$ $ ^{+0.039 }_{-0.043 }$}& 
{\scriptsize    0.283 $ ^{+0.053 }_{-0.031 }$ $ ^{+0.075 }_{-0.039 }$}&
{\scriptsize    9.07  $ ^{+2.78  }_{-0.73  }$ $ ^{+3.93  }_{-0.79  }$}&  
{\scriptsize $-$1.250 $ ^{+01.975}_{-04.246}$ $ ^{+02.395}_{-05.714}$}\\ 
\mr 
    {\scriptsize                    }&                                    
    {\scriptsize $h_0$              }& 
    {\scriptsize $q$                }&
    {\scriptsize $\chi^2_{\rm{min}}$}\\  
\mr
{\scriptsize $\Lambda$CDM                                        }& 
{\scriptsize    0.698 $ ^{+0.006}_{-0.006}$ $ ^{+0.009}_{-0.009}$}& 
{\scriptsize    0                                                }&                                                                                             
{\scriptsize    584.308                                          }&                                                     {\scriptsize                                                     }\\

{\scriptsize $\Lambda(t)$CDM                                     }& 
{\scriptsize    0.698 $ ^{+0.007}_{-0.007}$ $ ^{+0.010}_{-0.010}$}& 
{\scriptsize $-$0.198 $ ^{+0.528}_{-0.894}$ $ ^{+0.686}_{-1.235}$}& 
{\scriptsize    584.038                                          }&                      
{\scriptsize                                                     }\\                            
\br

\end{tabular}
\end{indented}
\end{table}

For the $\Lambda(t)$CDM model, the best-fit value of the interaction parameter indicates that the decay of dark energy into dark matter ($q<0$) is favoured over the inverse process. This result contrasts with that found in \cite{LtCDM} where the best-fit point (for a flat spacetime) was situated in the positive values of $q$. For $q>0$, the cosmological constant problem becomes actually worse, since the  dark energy density is  increasing with time and concerning the coincidence problem, as it is clearly shown from (\ref{eq:QLm}) and (\ref{eq:QLmk}), the interaction parameter must be negative in order to possibly have a solution with a constant  matter to dark energy density ratio at late times. In figure~\ref{fig:g15}, we  see that the presence of spatial curvature leads to an extension for the interval of confidence of each parameters relative to the flat case represented by the line $\Omega_{\rm k_0}=0$. In the case of the matter energy density $\Omega_{\rm m_0}$ and the radiation energy density $\Omega_{\rm r_0}$, this extension is clearly larger in the positive direction, while  for the interaction term $q$, it is in the negative direction. This latter is important since it allows for a large and negative value of $q$, which is more likely to solve the cosmological and the coincidence problems.

\begin{figure}[!ht]
  \centering
  \includegraphics[scale=0.650]{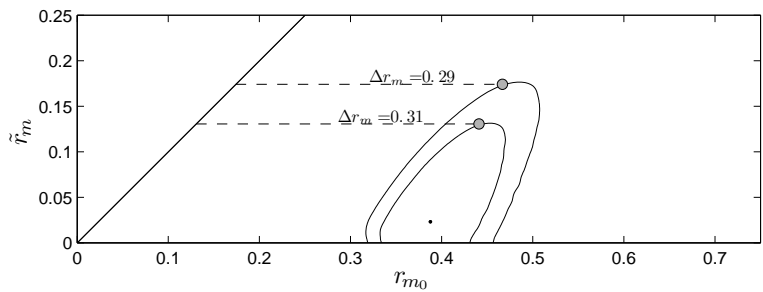} 
  \caption{Projections  of the 1-$\sigma$ and the 2-$\sigma$ confidence regions obtained from observational constraints  in the plane $r_{\rm m_0}-\tilde{r}_{\rm m}$. The parameter $\tilde{r}_{\rm m}$ is obtained from (\ref{eq:QLm}) and the parameter $r_{\rm m_0}$ is the current value of the matter to dark energy density ratio ($r_{\rm m_0}\equiv\Omega_{\rm m_0}/\Omega_{\rm \Lambda_0}$). The points minimizing the quantity $\Delta{r_{\rm m}}/r_{\rm m_0}$ in the 1-$\sigma$ and the 2-$\sigma$ confidence regions are indicated by the circular marks. The negative values of $\tilde{r}_{\rm m}$ which, inserted in (\ref{eq:QLm}), leads to a positive values of $Q_{\tilde{r}_{\rm m}}$ are not represented. Indeed, in this case $\tilde{r}_{\rm m}$ can be used as a parameter to characterize the decay of dark matter to dark energy but do not represent the physical ratio $\rho_{\rm m}/\rho_{\rm \Lambda}$ since we suppose that the interaction stops when all the dark matter has decayed ($r_{\rm m}=0$).}
  \label{fig:grm}
\end{figure}

In terms of the dimensionless parameter $q$, the critical value $Q_0$, which set the limit between the two classes of solutions (flat and non-flat)  becomes $q_0=-\sqrt{32\pi}\approx-10$. The largest (in terms of magnitude) negative value which lies the 2-$\sigma$ confidence region is $q=-1.433$; thus only the flat solutions are relevant here. We can then obtain the late-time value of $r_{\rm m}$ from the value of the interaction parameter by inverting (\ref{eq:QLm}), as we did in figure ~\ref{fig:grm}. In this figure, we have also shown the points, in the 1-$\sigma$ and  2-$\sigma$ confidence regions, which minimize the relative variation between  the current  and the late-time value of the matter to dark energy density ratio, $\Delta{r_{\rm m}}/r_{\rm m_0}=(r_{\rm m_0}-\tilde{r}_{\rm m})/r_{\rm m_0}$. Ideally, we would like to obtain $\Delta{r_{\rm m}}/r_{\rm m_0}\approx0$ in order to explain the coincidence problem (thus the current value of $r_{\rm m}$ would be typical for $t>t_0$). However, we get $\Delta{r_{\rm m}}/r_{\rm m_0}=0.70$ in the 1-$\sigma$ region and  $\Delta{r_{\rm m}}/r_{\rm m_0}=0.63$ in the 2-$\sigma$ region. Since  for the $\Lambda$CDM model,  $\Delta{r_{\rm m}}/r_{\rm m_0}\rightarrow1$, one can argue that the coincidence problem is alleviated in the $\Lambda(t)$CDM model. However, an interesting way to visualize the coincidence problem is to plot the function $F\equiv\min\left(r_{\rm m},r_{\rm m}^{-1}\right)$ as a function of the time (figure~\ref{fig:evol}). For the $\Lambda$CDM model (at the best-fit point), this function is characterized by an early and a late phase where $F\approx0$ which are separated by a median one where $F\napprox0$ and forms a peak.  The duration of this median phase is very narrow in comparison to the entire Universe history and the coincidence problem consists in the fact that we are currently situated in it. If we look now at the plot for the point minimizing $\Delta{r_{\rm m}}/r_{\rm m_0}$ in 2-$\sigma$ region in the $\Lambda(t)$CDM model, we can see that the function $F$ is characterized, as before, by an early phase where $F\approx0$ and a median one where $F\napprox0$ and forms a peak. However, for the late phase, $F\napprox0$ and becomes approximately constant ($F\rightarrow\tilde{r}_{\rm m}\approx0.17$). Since $r_{\rm m_0}\approx0.47$, the order of magnitude of the current value of $r_{\rm m}$ is now typical for $t>t_0$. In this sense the coincidence problem is alleviated. However, as was the case for the $\Lambda$CDM model, we are currently situated in the median phase, which remains narrow compared to the whole Universe history. Moreover, from the fluid evolution shown in figure~\ref{fig:evol}, we can see that at $t_0$, it is not only the energy densities of matter and of dark energy that are of the same order of magnitude, but also that of curvature, which is actually a triple coincidence problem. Hence, we can conclude that for the parameters range which are consistent with the observations, the $\Lambda(t)$CDM model fails to provide a completely satisfying explanation to the coincidence problem.

For both models, the discrepancy between the observed and the predicted values of the dark energy density remains roughly  the same  every everywhere in the 1 and the 2-$\sigma$ confidence regions ($\rho_{\Lambda_0}/\rho_{\Lambda_{th}}\sim10^{-123}$). For the $\Lambda(t)$CDM model, to explain this result, we have made the hypothesis that the dark energy density could decrease from an initial large value ($\rho_{\Lambda_i}$, evaluated at $a=0$), to the current observed one ($\rho_{\Lambda_0}$).  Inside of the 1-$\sigma$ confidence region, the largest value that we get for the ratio  $\rho_{\Lambda_i}/\rho_{\Lambda_{0}}$ is $\approx1.4$, and inside of the 2-$\sigma$ region, $\rho_{\Lambda_i}/\rho_{\Lambda_{0}}\approx 1.6$. These  are very far  from the ratio  $\rho_{\Lambda_i}/\rho_{\Lambda_{0}}\sim 10^{+123}$ needed to solve the cosmological constant problem. Hence, we can conclude that for the values of parameters that are consistent with the observations, $\Lambda(t)$CDM  is unable to provide an explanation to this problem.

Even if $\Lambda(t)$CDM fails to provide an explanation to both of the cosmological  problems, we notice that
the $\chi_{\rm min}^{2}$ values obtained for the interacting model (584.038) is slightly better than that obtained for the $\Lambda$CDM model (584.308). However, the interacting model involves an additional parameter to constrain ($q$), hence it is not surprising that it provides a better fit to data. To take into account the different number of parameters, the significance of the improvement of the $\chi^2_{\rm min}$ value may be assessed by the mean of the Bayesian information criterion \cite{BIC1}, defined as
\begin{equation}
   {\rm BIC}=-2\ln\mathcal{L}_{\rm max} + K\ln N=(\chi^{2}_{\rm min} + C) + K\ln N,
   \label{eq:BIC}
\end{equation}
where $\mathcal{L}_{\rm max}$ is the maximum likelihood, $K$ is the number of parameters for the model (4, for the $\Lambda$CDM model, 5 for the $\Lambda(t)$CDM), $N$ the number of data points used in the fit ($N=646$) and $C$ a constant independent of the model used. Following \cite{BIC2}, we will regard a difference of 2 for the BIC as a non-significant, and of 6 or more as very non-significant improvement of the $\chi^2_{\rm min}$ value. Using the $\Lambda$CDM model as reference, we get $\Delta_{\rm BIC}=6.2$.
Hence the addition of an extra parameter is not warranted by the marginal decrease in the value of $\chi^2_{\rm min}$. Since the $\Lambda(t)$CDM model is not able either to provide a satisfying explanation to the cosmological and to the coincidence problems we must conclude that the $\Lambda$CDM model remains the most satisfying one.

\section{Conclusion}
\label{sect:Conclusion}

Despite of successes (simplicity, good fit to data), the $\Lambda$CDM model is not completely satisfying because of the existence of the coincidence problem, and more importantly, of the cosmological constant problem. These two problems have been actively studied since the discovery of the accelerated expansion of the Universe, but none of the proposed solutions has been able to convince unanimously the cosmological community. In \cite{LtCDM}  a cosmological model where dark energy and dark matter interacts through a term $Q_{\rm \Lambda}=Q\rho_{\rm \Lambda}^{n}$ in flat spacetime was considered. It has been shown that for $n=3/2$, this model could have provided an elegant solution to both problems, but the values of the parameters required  were excluded by  observational constraints. 

Given the importance of the cosmological constant and the coincidence problem, we were motivated to complete the analysis of the model by checking whether this result holds also in the presence of spatial curvature. We have shown that even in that case, $n=3/2$ remains the only value for which it is possible to find a late-time cosmology with a constant ratio of dark matter to dark energy density. Depending on the strength of the interaction, the Universe will be nearly flat ($\rho_{\rm k}\approx0$) for $|Q|<|Q_0|$ or will admit spatial curvature ($\rho_{\rm k}\neq0$) for $|Q|>|Q_0|$ at late times. 

By constraining the model using observational data, we have found that within the 2-$\sigma$ confidence region, the cosmological constant problem remains  as severe as in the $\Lambda$CDM model. For the coincidence problem, the situation is different. It is now possible to find points in the 2-$\sigma$ confidence region for which the current value and the late-time value of the ratio of dark matter to dark energy density are of the same order of magnitude,  $\Or(r_{\rm m_0}/\tilde{r}_{\rm m})=1$. However, for these points, the current time $t_0$ is situated in the short time interval for which $r_{\rm m}\napprox0$ and $\dot{r}_{\rm m}\neq0$. Hence, we cannot conclude that, in presence of spatial curvature, the $\Lambda(t)$CDM model provides a completely satisfying solution to the coincidence problem.

\ack
We thank James Cline for his comments on this manuscript. This work was partly supported by the Fonds de recherche du Qu\'ebec - Nature et technologies (FQRNT) through the doctoral research scholarships programme.

\begin{appendices} 
\section{Observational constraints} 

To constrain the model parameters, we have proceeded similarly to what we did in \cite{LtCDM}, i.e. that we have used observational tests involving the distance modulus $\mu$ of type Ia supernova (SNeIa) and gamma-ray bursts (GRB), the baryon acoustic oscillation (BAO), the cosmic microwave background (CMB) and the observational Hubble rate (OHD). These data are actually frequently used to constrain the cosmological models with interacting dark energy \cite{cstr0,cstr1,cstr2,cstr4,cstr8,cstrA1,cstr3,cstr5,cstr6,cstr9}. In our current analysis, there are  two noticeable differences in comparison to our previous study. For the OHD constraints, we have used an updated dataset \cite{OHD0} and for the CMB constraints, we have only used the acoustic scale $l_{\rm A}$ since as it was pointed in \cite{cstr9}, the CMB shift parameter $R$ is  model dependent  and can be used only in the case where the dark energy density is negligible at the decoupling epoch (which a priori, we do not know). Moreover, it is to be noticed that the Planck results \cite{planck} came out after that we have started our numerical analysis. With these results, we could have used updated data for the CMB constraints ($\Omega_{\rm b_0}$ and $\Omega_{\rm \gamma_0}$) and for the BAO constraints, a measurement of the ratio  $r_{\rm s}(z_{\rm d})/D_{\rm V}(z)$ at a new redshift ($z=0.57$). For each of these dataset, the $\chi^2$ function is computed as
\begin{equation}
  \chi^{2}=\sum\limits_{i} \frac{[x_{\rm obs}(z_i)-x_{\rm th}(z_i)]^{2}}{\sigma_{i}^{2}},
  \label{eq:Chi}
\end{equation}
where $x_{\rm obs}$, $x_{\rm th}$ and $\sigma_{i}$ are respectively the observed value, the theoretical value and the $1$-$\sigma$ uncertainty associated with $i^{\rm th}$ data point of the dataset. The best fit is then obtained by minimizing the sum of all the $\chi^2$
\begin{equation}
   \chi^{2}_{\rm tot}=\chi^{2}_{\rm \mu}+\chi^{2}_{\rm OHD}+\chi^{2}_{\rm BAO}+\chi^{2}_{\rm CMB}.
   \label{eq:Xtot}
\end{equation}

\subsection{Distance modulus $\mu$ of SNeIa and GRB}
The distance modulus is the difference between the apparent magnitude $m$ and the absolute magnitude $M$ of an astronomical object. Its theoretical value for a flat Universe is given by
\begin{equation}
  \mu_{\rm th}(z)\equiv5\log_{10}\frac{D_{\rm L}(z)}{h_0}+42.38,
  \label{eq:mut}
\end{equation}
where $h_0\equiv H_{0}$/(100 ${\rm km}\ {\rm s}^{-1} {\rm Mpc}^{-1}$) and the Hubble free luminosity distance  is given by $ D_{\rm L}=(H_0/c)d_{\rm L}$. The luminosity distance, $d_{\rm L}$, is defined as
\begin{equation}
  d_{\rm L}(z)\equiv\frac{(1+z)}{(H_0/c)|\Omega_{\rm k_0}|^{\frac{1}{2}}}{\rm{sinn}}\left(|\Omega_{\rm k_0}|^{\frac{1}{2}}\int_{0}^{z}\frac{dz}{H(z)/H_0} \right).
\end{equation}
For a closed ($\Omega_{\rm k_0}<0$), a flat ($\Omega_{\rm k_0}=0$) and an open Universe ($\Omega_{\rm  k_0}>0$) the function ${\rm sinn }\ x$ is respectively equal to $\sin{x}$, $x$ and ${\rm sinh}\ x$. The observational data used are the 557 distance modulii of SNeIa assembled in the Union2 compilation \cite{SNE5} ($0.015 < z < 1.40$) and the 59 distance modulii of GRB from \cite{GRB} ($1.44 < z < 8.10$).  The combination of these two types of data covers a wide range of redshift  providing a more complete description of the cosmic evolution than the SNeIa data by themselves.

\subsection{Observational $H(z)$ data (OHD)}

The theoretical values of the Hubble parameter at different redshift is directly obtained from (\ref{eq:fried1}).
We have 
The observational data have been taken from \cite{OHD0} where 28 values (ranging from a redshift of $z=0.070$ to $z=2.30$) have been compiled.

\subsection{Baryon acoustic oscillation  (BAO)}
The use of BAO to test a model with interacting dark energy is usually made \cite{cstr0,cstr1,cstr2,cstr4,cstr8,cstrA1,cstr3,cstr5,cstr6,cstr9} by means of the the dilation scale
\begin{equation}
  D_{\rm V}(z)=\left[\frac{z(1+z)^2}{H(z)/c}d_{\rm A}^{2}(z)  \right]^{1/3}.
\end{equation}
Since the (proper) angular diameter distance $d_{\rm A}$ is related to the luminosity distance $d_{\rm L}$ through $d_{\rm A}=d_{\rm L}/(1+z)^2$, the dilation scale may also be expressed as
\begin{equation}
  D_{\rm V}(z)=\left[\frac{z}{(H(z)/c)(1+z)^2}d_{\rm L}^{2}(z)  \right]^{1/3}.
\end{equation}
The ratio  $r_{\rm s}(z_{\rm d})/D_{\rm V}(z)$, where $r_{\rm s}(z_{\rm d})$ is the comoving sound horizon size at the drag epoch,  has been observed at $z=0.35$ by SDSS \cite{Einsenstein} and at $z=0.20$ by 2dFGRS \cite{DV_ratio}. To avoid to have to compute $r_{\rm s}(z_{\rm d})$, we will minimize the $\chi^2$ of the ratio $D_{\rm V_{0.35}}/D_{\rm V_{0.20}}$. The observed value for this ratio is $1.736\pm0.065$ \cite{DV_ratio}.

\subsection{Cosmic Microwave Background (CMB)}
\label{sec:CMB}

The values extracted from the 7-year WMAP data for the acoustic scale ($l_{\rm A}$) at the decoupling epoch ($z_*$) can be used to constrain the model parameters ($l_{\rm A}(z_*)=302.09\pm0.76$). Its theoretical value is computed as \cite{CMB3}
\begin{equation}
  l_{\rm A}(z_*)=\pi(1+z_*)\frac{d_{\rm A}(z_*)}{r_{\rm s}(z_*)}
\end{equation}
where $r_{\rm s}(z_*)$, the comoving sound horizon at the decoupling epoch, is given by
\begin{equation}
  r_{\rm s}(z_*)=\int^{\infty}_{z_*}c_{\rm s}\frac{dz}{H}.
\end{equation}
Hence, in term of the luminosity distance,  the acoustic scale may be expressed as
\begin{equation}
  l_{\rm A}(z_*)=\frac{\pi}{(1+z_*)}\frac{d_{\rm L}(z_*)}{\int^{\infty}_{z_*}c_{\rm s}\frac{dz}{H}}.
\end{equation}
In these expressions, the  sound velocity is given by 
\begin{equation}
  c_{\rm s}=c\left(3+\frac{9}{4}\frac{\rm \Omega_{b_0}}{\Omega_{\gamma_0}(1+z)}\right)^{-1/2}, 
\end{equation}
and following \cite{cstr1,cstr2,cstr4,cstr8,cstrA1}, we use the following fitting formula \cite{fitting} to find $z_*$
\begin{equation}
   z_*=1048[1+0.00124(\Omega_{\rm b_0} h_0^2)^{-0.738}][1+g_1(\Omega_{\rm m_0} h_0^2)^{g_2}],
   \label{eq:fitz}
\end{equation}
where
\begin{eqnarray}
g_1 & \equiv0.0783(\Omega_{\rm b_0} h_0^2)^{-0.238}(1+39.5(\Omega_{\rm b_0} h_0^2)^{-0.763})^{-1},\\
g_2 & \equiv0.560(1+21.1(\Omega_{\rm b_0} h_0^2)^{1.81})^{-1}.
\end{eqnarray}
Two additional parameters are needed to determine the acoustic scale, namely the current value of the density parameter of baryons ($\Omega_{\rm b_0}$) and of radiation ($\Omega_{\rm \gamma_0}$). Constraining the model with these two additional parameters will require in an increased computational cost. However as suggested in \cite{cstr0}, we can use the values obtained in the context of the $\Lambda$CDM cosmology. This is motivated since the radiation and the baryons are separately conserved, and because we want to preserve the spectrum profile as well the nucleosynthesis constraints. The observational results from 7-year WMAP data \cite{CMB4} are 
\begin{equation}
  \Omega_{\rm b_{0}}=2.25\times 10^{-2}h_0^{-2} \quad {\rm and } \quad \Omega_{\rm \gamma_{0}}=2.469\times 10^{-5}h_0^{-2}.
\label{eq:Ob0}
\end{equation}
These two quantities are related to two of the constrained parameters since ${\Omega_{\rm m_{0}}=\Omega_{\rm dm_{0}}+\Omega_{\rm b_{0}}}$ and $\Omega_{\rm r_{0}}=\left( 1+7/8(4/11)^{4/3}N_{\rm eff}\right) \Omega_{\rm \gamma_{0}}$. The combinations of initial parameters $\Omega_{\rm m_0}$, $\Omega_{\rm r_0}$ and $h_0$ which lead to a negative energy density for dark matter ($\Omega_{\rm dm_{0}}<0$) or to an effective number of neutrinos species smaller than three ($N_{\rm eff}<3$) have been excluded from our analysis.

\end{appendices}

\end{document}